**The Copernican Argument for Alien Consciousness;**
**The Mimicry Argument Against Robot Consciousness**

Eric Schwitzgebel
Department of Philosophy
University of California, Riverside
Riverside, CA 92521
USA

Jeremy Pober
Centre of Philosophy, University of Lisbon
1649-004 Lisboa, PT

26 Jan 2026

**The Copernican Argument for Alien Consciousness;**
**The Mimicry Argument Against Robot Consciousness**

Abstract: On broadly Copernican grounds, we are entitled to assume that apparently behaviorally sophisticated extraterrestrial entities ("aliens") would be conscious. Otherwise, we humans would be inexplicably, implausibly lucky to have consciousness, while similarly behaviorally sophisticated entities elsewhere would be mere shells, devoid of consciousness. However, this Copernican default assumption is canceled in the case of behaviorally sophisticated entities designed to mimic superficial features associated with consciousness ("consciousness mimics"), and in particular a broad class of current, near-future, and hypothetical robots. These considerations, which we formulate, respectively, as the Copernican and Mimicry Arguments, jointly defeat an otherwise potentially attractive parity principle, according to which we should apply the same types of behavioral or cognitive tests to aliens and robots, attributing or denying consciousness similarly to the extent they perform similarly. Our approach is unusual in the following respect: Instead of grounding speculations about alien and robot consciousness in a particular metaphysical or scientific theory about the physical or functional bases of consciousness, we appeal directly to the epistemic principles of Copernican mediocrity and inference to the best explanation.

**The Copernican Argument for Alien Consciousness;**
**The Mimicry Argument Against Robot Consciousness**

*1. Our Argument in Sketch.*

Engineers might soon build robots so sophisticated that people are tempted to regard them as meaningfully conscious or sentient – robots capable, or seemingly capable, of genuinely feeling pleasure and genuinely suffering; robots that possess, or seemingly possess, real conscious awareness of themselves and their environments.[1] Less likely, we might soon discover complex, seemingly intelligent, seemingly conscious space aliens – maybe beneath the ice of Europa or orbiting a nearby star. It would be lovely if we had a scientifically well-justified consensus theory of consciousness that we could straightforwardly apply to determine whether such robots or aliens are genuinely conscious or only superficially seem so. Unfortunately, a consensus theory is unlikely anytime soon.[2]

Still, as we will argue, we are not wholly in the dark regarding robot and alien consciousness. When confronted with what we call *behaviorally sophisticated* robots or aliens, a particular pair of default attitudes is justified.

Toward behaviorally sophisticated extraterrestrial entities ("aliens"), the appropriate default is to attribute consciousness. We will defend this view on Copernican grounds.

---

[1] By "consciousness" we mean "phenomenal consciousness" as standardly used in mainstream Anglophone philosophy of mind, innocent of any commitment to naturalistically dubious properties such as immateriality or irreducibility (see discussion in Schwitzgebel 2016). Researchers who regard near-future robot consciousness as a real possibility include Dehaene, Lau, and Kouider 2017; Chalmers 2023; Butlin et al. 2023. Dissenters include Godfrey-Smith 2016; Block 2025; Seth forthcoming.

[2] Seth and Bayne 2022 survey recent scientific theories of consciousness. Approaches relying on points of agreement among competing theories, such those advocating "ecumenical heuristics" (Shevlin 2020), "theory light" approaches (Birch 2022), or "indicator properties" (Butlin et al. 2023), though helpful, face tradeoffs between substance and breadth of appeal.

According to the cosmological principle of Copernican mediocrity, it's reasonable to assume as a default that we do not occupy any special position in the universe. Copernican mediocrity would be violated if, among all the behaviorally sophisticated entities that presumably exist in the universe, we humans are among the small proportion who also happen to be conscious, while the rest are, so to speak, empty shells. That would make us weirdly, inexplicably lucky – much as if we just happened, inexplicably, to occupy the exact center of the universe.

Toward a broad range of behaviorally sophisticated current, near-future, and hypothetical robots – entities we call "consciousness mimics" – we will argue that the appropriate default is skepticism or uncertainty about their consciousness. If a system is designed to mimic the behavior (for example, text outputs) of conscious entities to evoke a reaction in an audience, we normally cannot justifiably infer the presence of consciousness from features that would otherwise suggest it. Further argument is needed – and in our current scientific and philosophical context, all such further arguments are dubious.[3] There's a vast gulf between a robot designed or selected to emit the superficial appearance of a friendly greeting – a robot perhaps modeled on us and engineered specifically to pass various tests – and a robot who actually experiences friendship.

Thus, the alien and robot cases are asymmetric.[4] If an alien exits a spaceship and greets you, it's reasonable to assume that you have encountered a conscious being; if a social robot designed to mimic human behavior rolls off a factory floor and greets you, it's reasonable to refrain from that assumption. This asymmetry holds even if the alien and robot exhibit an overall similar degree of behavioral sophistication.

[3] For a review, see Schwitzgebel 2025/2026.

[4] As we hope will become clear, the Copernican Principle would apply to a terrestrial 'alien' species that had gone unnoticed by humanity for our entire existence (such as the Silurians from Doctor Who). And the mimicry argument would apply to an extraterrestrial, biological consciousness mimic (see Section 10).

A certain type of parity principle might have seemed plausible, according to which we should approach robots and aliens similarly, applying the same types of behavioral or cognitive tests and attributing or denying consciousness similarly if they perform similarly. We disagree. As we will argue, the history of an entity's creation makes an epistemic difference to our assessment of its consciousness, independent of its behavioral or cognitive performance.

## *2. Behavioral Sophistication.*

The concept of *behavioral sophistication* is central to our framework. An entity is "behaviorally sophisticated" if it is capable of complex goal-seeking, complex communication, and complex cooperation, where these terms describe behavioral patterns without presupposing consciousness or any particular internal cognitive architecture.

Suppose an alien species constructs intricate devices that extract nutrition from its environment. These nutrition sources are stored and consumed by conspecifics as needed. Individuals respond to environmental threats with highly specific evasive activities, often well in advance and with narrow margins for error. The aliens interact in highly complex and seemingly technologically advanced ways, engaging in activities that increase their chances of survival or reproduction only if several others engage in specific activities at specific times while physically remote and not in direct contact. This interaction is enabled by complex signals with flexible contents. For good measure, suppose that the aliens fly spacecraft at least as good as ours.

Such a species would be *behaviorally sophisticated* as we intend the phrase. When biologists discuss the goals, communication, and cooperation of bacteria, birch trees, and mycorrhizal fungi, such terms need not presuppose consciousness or cognitive mechanisms familiar from animal cases. Similarly for our use of "goals", "communication", and "cooperation" here. (Employ scare quotes if you like.) Perhaps behavioral sophistication in our sense metaphysically requires consciousness or a particular cognitive architecture. But that depends on theories among which we aim to stay neutral. The point we will insist on is that behavioral sophistication bears an *epistemic* relationship to consciousness.

How sophisticated is "behaviorally sophisticated"? Our Copernican Argument works in its simplest form if approximately human-level sophistication is the required minimum. However, as we suggest in Section 14, the Copernican Argument can be generalized to a much lower standard of sophistication. To forestall a possible confusion: Our argument will be that human-level behavioral sophistication is a *sufficient* condition for the default attribution of consciousness, not a necessary condition.

*3. The Copernican Principle of Consciousness.*

We assume that behavioral sophistication is not *extremely* rare – that it has evolved or will independently evolve at least a thousand times in the observable universe (roughly once per billion galaxies). The true number is plausibly orders of magnitude greater.[5]

Among these many instances of behavioral sophistication, the distribution of consciousness could vary dramatically. At one extreme, all behaviorally sophisticated extraterrestrial ("alien") entities might be conscious. At the other extreme, only Earthlings are. A vast range lies between. If among all these entity-types, only we Earthlings were conscious, or only we and a few others, we would be strikingly special. According to an epistemic principle we call the Copernican Principle of Consciousness, we can default assume that we aren't in that way special.

[5] For scientific estimates of the frequency of "intelligent" or "technological" life in the universe, see Frank and Sullivan 2016; Snyder-Beattie et al. 2021.

The Copernican Principle of Consciousness is a special case of the Copernican Principle in cosmology, according to which "the Earth is not in a central, specially favored position" (Bondi 1968, p. 13) or, alternatively, "We do not occupy a privileged position in the Universe" (Barrow and Tipler 1986, p. 1; various formulations are possible, e.g., Scharf 2014).[6] Recent scientific applications tend to emphasize the principle's connection to the homogeneity and isotropy of the universe at large scales.[7] If everything looks pretty much the same everywhere, no position – including ours – is particularly special.

Extending this thinking to consciousness, we propose the *Copernican Principle of Consciousness:*

> Among behaviorally sophisticated entities, we are not specially privileged with respect to consciousness.

If the Copernican Principle of Consciousness is correct, we are not specially lucky to possess consciousness-instilling Earthiform biology while other behaviorally sophisticated entities' architectures leave them entirely nonconscious. Nor are we especially *more* conscious than other behaviorally sophisticated entities or blessed with a special kind of awesomely *good* consciousness. Among all spaceship-flying, linguistically communicating, long-term-goal planning, socially cooperative entity-types in the universe, only we (and maybe a few others) are conscious?! Although that *could* be so, we should not assume that we're uniquely bestowed with light. Absent some reason to think we are special, we Earthlings would then be suspiciously, inexplicably lucky.

[6] We use the formulation that *we* are in a special position, rather than the alternate formulation that *Earth* is, since it more clearly displays that it's a self-location principle.

[7] E.g., Caldwell and Stebbins 2008; Clarkson, Bassett, and Lu 2008; Camarena, Marra, Sakr, and Clarkson 2022.

Copernican principles are *default* epistemic principles. Compelling evidence could potentially arise that we *are* at the center of the universe. Maybe our best theory of consciousness will someday imply we're lucky to have Earthiform neurons instead of Andromedean hydraulics. The assumption of mediocrity is just a reasonable starting place.

### *4. Three Concerns about the Copernican Principle of Consciousness.*

We hope that most readers will find the Copernican Principle of Consciousness plausible on its face. However, to clarify and assess its commitments, we consider three concerns.

*Concern A. "Specialness" is vacuous.* Copernican principles depend on the somewhat murky notion of "special privilege".[8] Specified narrowly enough, any position could be identified as unique or extremely unusual on some metric of evaluation. Maybe Earth is the only planet in the observable universe with a mass of X, average orbital distance of Y, and moon with magnesium-to-iron ratio of Z, with X, Y, and Z specified to 100 digits. Similarly, if I deal five cards and receive J♣, 4♦, 7♣, Q♠, 7♠ in that order, in some sense that is a "special" or low-probability outcome, since the chance of receiving those exact cards in that order is under one in three hundred million. But if everything is special, then claims of specialness are nearly vacuous and Copernican reasoning collapses.

[8] Partly as a result, they invite concerns similar to those of other "self-location" principles, such as the Doomsday Argument (Gott 1993; Lacki 2023), the Sleeping Beauty puzzle (Elga 2000 and subsequent literature), and anthropic principles (more on which below) (Barrow and Tipler 1986; Bostrom 2002; and subsequent literature).

Responding to this worry requires clarifying “special”. In poker, A♠, K♠, Q♠, J♠, 10♠ is a special or privileged hand, because no other hand can beat it. For poker purposes, a royal flush has distinctive importance. But even outside poker, A♠, K♠, Q♠, J♠, 10♠ is striking in a way J♣, 4♦, 7♣, Q♠, 7♠ is not (unless something hinges on drawing those exact cards). Similarly, occupying the precise center of the universe is intuitively striking in a way that being 3.24 billion light years from the center of an approximately Hubble-sized universe is not. A license plate of OOO 000 is intuitively striking in a way that XNE 488 is not. Seeing OOO 000 on your neighbor’s new car, you can reasonably expect an explanation other than chance.[9] Alternatively, specialness might be captured by information compressibility: A royal flush can be described as “the top five spades in descending order”, while describing the other hand requires listing the specific cards. We can describe the center of the universe as just “the center”, while most other positions require three spatial coordinates.[10] Thus, “specialness” can be characterized in at least three ways: relative to goals or rewards, in terms of something like intuitive strikingness, and in terms of information compressibility. Underlying all three is the idea that “specialness” requires more than simple rarity. If any of these three approaches works, Copernican principles are not in general vacuous.[11]

---

[9] For a recent discussion of surprisingness and strikingness in philosophy of probability, see Baras 2022.

[10] One challenge for compressibility concerns variation in language: The mixed hand would be highly compressible in language with a one-bit representation for J♣, 4♦, 7♣, Q♠, 7♠ (1) vs. all other distributions (0). A fully developed compressibility view requires either objective measures of compressibility (perhaps in terms of negative entropy) or privileged (natural kind?) categories.

[11] Another approach (Gott 1993) might drop the concept of “specialness” entirely, substituting the idea that we shouldn’t expect to find ourselves near the extremes of *any* property, special or not, unless we’ve cheated by either choosing a property with respect to which we already know ourselves to be extreme or checking many properties (thus facing the statistical issue of multiple comparisons).

*Concern B. Consciousness is not special.* The Copernican Principle of Consciousness requires specifically that *consciousness* is special – that being the one conscious entity in a pool of behaviorally sophisticated aliens would be more like drawing A♠, K♠, Q♠, J♠, 10♠ than like drawing J♣, 4♦, 7♣, Q♠, 7♠. But maybe consciousness isn't special?

On its face, consciousness is extraordinarily special. Some philosophers regard it as the most intrinsically valuable thing in the universe, the grounds of moral status, and the primary source of meaning in life.[12] Even if we don't go that far, that consciousness is even a plausible candidate for such high status suggests its importance. If you were to learn you would never again be conscious, you would probably be dismayed in a way you would not be dismayed to learn you will never again weigh exactly X grams. Consciousness is also an intuitively striking property: If we somehow discovered that we human beings are conscious while our nearest behaviorally sophisticated alien neighbors are not, that would be a remarkable difference. Arguably, too, when we describe an entity as "conscious", we compress substantial information into a brief statement, assuming that the attribution of consciousness captures some general pattern or capacity in the target entity – though exactly which pattern or capacity will depend on what account of consciousness is correct or on what patterns or capacities the attributer regards as characteristic.[13]

Our Copernican principle doesn't require consciousness to be *uniquely* special in the manner suggested by strong statements of its specialness; and, if necessary, it could be adapted to reference some other nearby property (e.g., illusionist "quasi-consciousness"). One might also potentially argue that although consciousness is special in a way, we should expect behaviorally intelligent alien species to have other properties that are equally special – schmonsciousness, say, for the aliens in Galaxy Alpha, bronsciousness for those in Galaxy Beta, and lulonsciousness for those in Galaxy Gamma which nonetheless equally give their lives meaning and value (Lee 2019). We could then relativize the Copernican Principle to the disjunction of schmonsciousness-or-bronsciousness-or-lulonsciousness-etc. or to whatever generic property those specific properties share.

[12] Discussions include Kriegel 2019; Shepherd 2024. For more skeptical perspectives, see Levy 2024; Papineau 2024; as well as "illusionists" about consciousness such as Frankish 2016 and Kammerer 2019.

[13] We thank Nick Shea for emphasizing this issue in conversation.

*Concern C. Reference groups and the Anthropic Principle.* It might be objected that our principle depends on an arbitrary choice of reference class. Suppose, for example, we stated the principle as: Among entities containing carbon, we are not specially privileged with respect to consciousness. Such a principle would fail, since within that class – which includes plants, lumps of coal, and late-stage carbon-burning stars – we *are* specially privileged with respect to consciousness. What justifies taking behaviorally sophisticated entities as the reference group with respect to which we assume we aren't specially privileged?[14]

Our answer: Copernican reasoning only applies absent countervailing evidence. And unless the mainstream scientific worldview is wildly wrong, we already know that our degree of consciousness is special among the class of entities containing carbon; so that's not an appropriate reference class. By focusing on behavioral sophistication, we target a question on which we lack direct evidence: whether very differently constructed but behaviorally sophisticated aliens would be conscious.

The original Copernican Principle faces a similar reference group challenge. We *are,* after all, in a special location: a rare spot of the cosmos hospitable to complex life (e.g., not vacuum, not inside a star, a location rich with complex molecules in a good solvent). To address this challenge, cosmologists can add an *anthropic* qualification. As Brandon Carter remarks, our position is "necessarily privileged to the extent of being compatible with our existence as observers" (Carter 1974, p. 293; see also Barrow and Tipler 1986; Bostrom 2002). We are not in a special location relative to other locations *capable of supporting observers*. Can we add a similar anthropic qualification to the Copernican Principle of Consciousness?

[14] We thank Andrew Lee for emphasizing this issue in conversation.

That depends on what counts as an "observer". If observers must be conscious, the anthropically qualified Copernican Principle of Consciousness becomes: "Among behaviorally sophisticated entities *who are conscious observers*, we are not specially privileged with respect to consciousness." This principle isn't vacuous, since it implies that we don't have especially more or better consciousness. But the principle is useless for our purposes, since it doesn't address the question of whether behaviorally sophisticated aliens in general are conscious. The nonconscious ones would be ruled out of the comparison class.

Alternatively "observer" might be operationalized behaviorally, so that behavioral sophistication alone qualifies an entity as an "observer" for purposes of the Copernican Principle of Consciousness. In that case, our original Copernican Principle of Consciousness already contains the required anthropic qualification: "Among behaviorally sophisticated entities [i.e., "observers"], we are not specially privileged with respect to consciousness." Thus, we have two ways to meet the reference-class challenge: First, treat the Copernican Principle as applying only when we stand in sufficient ignorance about the consciousness of entities in the reference class. Second, treat the Copernican Principle as already implicitly anthropically constrained to behaviorally operationalized "observers".

*General Observations.* All three concerns arise from worries about applying Copernican reasoning too broadly. Pick any property you like – ten-fingeredness, say – and any reference class you like – entities weighing, say, between 5 and 500 kilograms – and unrestrained Copernican reasoning will absurdly deliver surprise that there aren't more ten-fingered boulders. Now, our Copernican reasoning *is* intended very generally. We prefer not to restrict it a priori to any particular domain. But as we hope our replies make clear, good Copernican reasoning is nonetheless severely restricted, both in target property and in reference class. The target properties must be few and special. If ten-fingeredness works, presumably so also do two-armedness, brown-skinnedness, music-likingness, and maybe also inhabiting a planet with a country shaped like a boot. Unlike these properties, and like being at the exact center of the cosmos, consciousness has a specialness that warrants an inference of mediocrity. And once we choose consciousness as our target property, the reference class cannot be any group that we have good independent reason to think differs from us with respect to consciousness. This constraint excludes almost all possible reference classes, though it includes the reference class of behaviorally sophisticated entities (or possibly, as we will discuss in Section 11, behaviorally sophisticated entities that are not consciousness mimics).

Here's another way to capture the spirit of the Copernican Principle of Consciousness: Imagine a third alien entity (perhaps conscious) who observes human beings and other behaviorally sophisticated entities in other locations. This observing alien should presumably see us as just one of the bunch. This neutral – not godlike, but not Earth-centered – perspective is the proper Copernican perspective. We shouldn't regard our consciousness as special, unless there's some evidence that we are special that could be recognized from an unbiased outside perspective. Otherwise we err like the self-absorbed teen who regards their adolescent anguish and yearnings as somehow invisibly unique.

We hope the reader finds the fundamental idea plausible. There would be something odd, something strikingly un-Copernican, in supposing that among all the thousands or billions or quadrillions of behaviorally sophisticated entities in the trillion galaxies we can see, only we, or only we and a small proportion of others, have the good fortune to be conscious; all the rest merely *seem* to feel love, to experience ecstasies of poetry, to share excited news of their discoveries, and to thrill with reverent awe before towers to their gods, while being as experientially empty as toasters.

### *5. From the Copernican Principle to Default Liberalism about Consciousness.*

The Copernican Principle of Consciousness holds that we are not *specially* privileged with respect to consciousness. It doesn't follow that we aren't *slightly* privileged. A royal flush in a single draw rightly draws suspicions of funny business, but two pairs could easily be chance. Maybe only 10% of behaviorally sophisticated entities are conscious? Then we wouldn't have to be *too* lucky to be among them.

Let's define *Default Liberalism* about consciousness as the view that, absent specific grounds for doubt, we are warranted in attributing consciousness to any behaviorally sophisticated entity. Symmetry and simplicity justify moving from the Copernican Principle of Consciousness to Default Liberalism. If some behaviorally sophisticated entities are nonconscious, that shouldn't be just an arbitrary fact about them. We should expect some striking difference between conscious and nonconscious aliens that constitutes specific grounds for doubt.[15]

Suppose we discover an otherwise apparently behaviorally sophisticated alien species that lacks behavior suggestive of complex self-representation. Or suppose we discover a species that appears to lack flexible learning. Even independently of any specific general theory of consciousness, such striking general deficits might warrant concerns about corresponding deficits in consciousness.[16] Our approach can handle these cases in one of two ways. We could define "behavioral sophistication" to exclude such entities: Behavior doesn't count as sophisticated unless it meets certain specific standards (standards suggestive of complex self-representation and flexible learning). Alternatively, we could liberally describe such entities behaviorally sophisticated overall if they are advanced enough in other ways. Doubt is justifiable either way: either because they are not "behaviorally sophisticated" or because, even though they are, the difference is plausibly important enough to cancel the default.

We won't attempt to specify precisely what deficits would justify withholding the attribution of consciousness. One approach would be to focus on features highlighted by influential theories of consciousness. Internal structural anomalies might also cancel the default, such as if we discovered that the aliens were (somehow) entirely empty inside or made of pure, undifferentiated wood.

[15] We bracket views on which the presence of consciousness correlates with nothing of functional, or otherwise striking, importance. Notably, even anti-functionalists such as Chalmers (1996) and Block (2025) expect that in naturally evolved entities, consciousness will correlate with something functionally important. Since our claim concerns only natural regularities, not metaphysical necessities, functionally identical but naturally impossible "zombies" are irrelevant to the argument.

[16] On some theories of consciousness ("higher-order" theories and "unlimited associative learning" theories, respectively), self-representation and flexible learning are crucial to consciousness. For a review of higher-order theories, see Carruthers and Gennaro 2001/2023. On Unlimited Associative Learning, see Ginsburg and Jablonka 2019.

In sum, Copernican considerations justify thinking that among behaviorally sophisticated entities we are not highly unusual in being conscious. Simplicity or symmetry then further suggests that any division between conscious and nonconscious behaviorally sophisticated entities should correlate with some important difference between them. Both principles generate only default attributions that can be canceled given reasonable grounds for doubt.

Our argument is not: Sophisticated behavior entails the right functional roles or internal structures for consciousness (according to some particular functionalist or structuralist theory). Rather it is: Sophisticated behavior warrants a default attribution of consciousness (per Copernican mediocrity and simplicity or symmetry), with the default cancelable by any of a wide range of plausible counter-considerations.

## *6. Mimicry Arguments Against Robot Consciousness.*

We now turn to what we call Mimicry Arguments against Robot Consciousness. In the context of Copernican Default Liberalism about consciousness, Mimicry Arguments constitute specific grounds for doubt about the consciousness of a particular class of entities: *consciousness mimics*. We can be Copernican liberals about behaviorally sophisticated entities in general while reserving skepticism about mimics.

AI consciousness skeptics commonly warn against mistaking imitations of consciousness for the real thing. Emily Bender and colleagues (2021) characterize Large Language Models, like ChatGPT, as "stochastic parrots" that copy the form of language with no understanding of its meaning. Johannes Jaeger (2023/2024) recommends replacing the phrase "artificial intelligence" with “algorithmic mimicry” to emphasize that AI systems merely mimic intelligence without possessing it. John Searle argues that computers only “simulate” mental processes, and “no one supposes that a computer simulation of a storm will leave us all wet” (1984, p. 37-38). Kristin Andrews and Jonathan Birch (2023) warn us against too readily attributing sentience to AI systems designed specifically to mimic human responses. Daniel Dennett (2023) warns against language models as “counterfeit people”.

Some of the best-known philosophical thought experiments meant to cast doubt on AI consciousness implicitly or explicitly involve mimicry. In Searle’s (1980) “Chinese room”, a person who knows no Chinese but who has an extremely large rulebook is hypothesized to interact with the outside world indistinguishably from a real Chinese speaker by accepting Chinese characters as inputs and issuing other Chinese characters as outputs in accord with the rulebook’s advice. Block’s (1978/2007) “Chinese nation” thought experiment (relatedly, the “Blockheads” of Block 1981) imagines a large population controlling a body by consulting a bulletin board in the sky and following simple if-then rules to duplicate the functional structure of a real human being for an hour. Without entering into the details of these thought experiments or assessing their persuasiveness, we note a common feature: The systems are specifically designed to imitate human behavior.

Mimicry arguments can at least sometimes be excellent grounds for doubt about the consciousness of an AI system, for the same reason that we might reasonably doubt that an online tech support agent is genuinely pleased to help if we learn that it’s a chatbot imitating the superficial signs of enthusiastic helpfulness. However, advocates of mimicry arguments have not generally contextualized their arguments in a theory of mimicry that explains why this is so.

The remainder of this essay addresses this challenge, then explores how to reconcile mimicry-based concerns about AI consciousness with our Copernican Principle.

*7. Mimicry in General.*

Suppose you encounter something that looks like a rattlesnake. One possible explanation is that it is a rattlesnake. Another is that it *mimics* a rattlesnake. Such mimicry can arise through evolution (other snake species that evolve the superficial appearance of a rattlesnake to discourage predators) or through human design (a rubber rattlesnake). Normally, it's reasonable to suppose that things are as they seem. But this default can be overturned – for example, if there's reason to suspect frequent mimics.

In biology, *deceptive mimicry* (for example, Batesian mimicry [Bates 1862/1981]) occurs when one species (the mimic) resembles another species or entity (the model) in order to mislead another species such as a predator (the receiver). For example, viceroy butterflies evolved to resemble monarch butterflies, exploiting predators' aversion to the monarch's toxicity.[17] Similarly, gopher snakes evolved to shake their tails in dry brush, resembling the look and sound of rattlesnakes. Chameleons and cephalopods mimic their physical environment to mislead predators into treating them as unremarkable continuations of that environment.

*Social mimicry* occurs when one animal behaves similarly to another for social advantage. For example, African grey parrots imitate each other to facilitate bonding and signal in-group membership, and their imitation of human speech arguably functions to increase the affection of human caretakers. In deceptive mimicry, the superficially imitated feature normally does not correspond with the model's relevant trait. The viceroy is not toxic, and the gopher snake has no poisonous bite. Social mimicry need not have a deceptive purpose, but even so, the signal might or might not correspond with the underlying trait: the bird might or might not belong to the group it is imitating, and Polly the parrot might or might not really "want a cracker".[18]

[17] The viceroy case, though standard, is probably more complex than this simple portrayal: Prudic, Timmermann, Papaj, Ritland, and Oliver 2019.

All mimicry thus involves three traits: the readily observable trait (S2) of the mimic, the corresponding observable trait (S1) of the model, and a further feature (F) of the model that is normally indicated by the presence of S1 in the model. All mimicry also involves three protagonists: a model, a mimic, and a receiver whose reactions to S2 explain the resemblance of S2 to S1. Crucially, in mimicry, the resemblance of S2 to S1 is not directly explained by the presence of F in the mimic. Instead, it is explained by a potential receiver's reaction to that resemblance, given that the receiver treats S1 as normally indicating F. See FIGURE 1.

[18] For reviews of the biology of mimicry, see Jamie 2017; Anderson and de Jager 2020.

FIGURE 1: The mimic's possession of a readily observable feature S2 is explained by its resemblance to observable feature S1 in the model, because of how a receiver, who treats S1 as indicating F, responds to the resemblance of S1 and S2. S1 reliably indicates F in the model, but S2 need not reliably indicate F in the mimic.

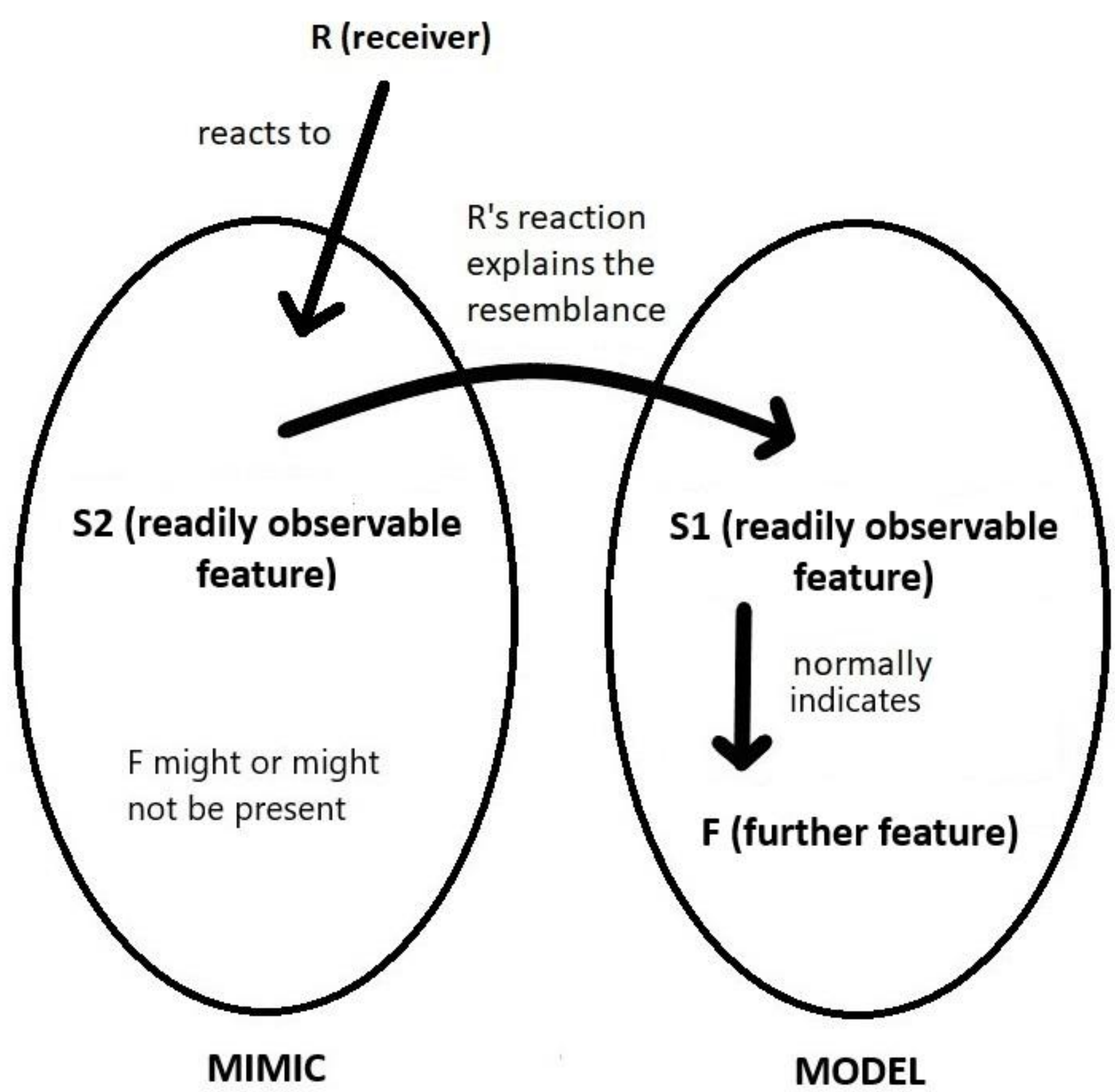

Six clarifications:

(a.) The protagonists need not be distinct. In parrots' social mimicry, the model might also be the receiver. If you mimic someone merely to amuse yourself, then you are both mimic and receiver.

(b.) The receiver need not be fooled. The parrot owner doesn't mistake Polly's speech for human speech. In the wild, the parrot does not mistake its mate's singing for its own.

(c.) The receiver's reaction to S1 and S2 need not be similar. A parrot might react very differently to a mate's imitation than to its own song. A plaza clown mimicking a depressed person's gait might elicit laughter rather than sympathy. What matters isn't sameness of response but only that the resemblance of S2 to S1 be explained by the reactions of potential receivers, given that the receivers treat S1 as an indicator of F.

(d.) The mimic need not lack F. The viceroy might happen to be toxic to some predator species. As long as S2 in the mimic is better explained by its resemblance to S1 than by the presence of F in the mimic, it's still mimicry.

(e.) Emulation or imitation alone is insufficient for mimicry. One animal might imitate another for a variety of reasons, for example, to duplicate their success in a challenging task. Mimicry distinctively requires a specific relationship to a receiver.[19]

(f.) Mimicry involves a gap between S1 and F. If I imitate your holding up three fingers by also holding up three fingers, I have copied or imitated you, but I haven't necessarily mimicked you. If the three fingers are a signal of membership in the Cool Kids Club and I raise them in imitation of you to trick the doorkeeper, then I have mimicked you.[20]

---

[19] On emulation and imitation, see Whiten et al. 2009.

[20] Readers of evolutionary accounts of mimicry might wonder whether "honest" mimicry fits into the present account. In evolutionary biology, Müllerian mimicry occurs when a toxic species mimics the pattern of another toxic species to more effectively signal toxicity to predators who can learn the signal more efficiently given the similarity of mimic and model

Mimicry is thus a specific, causally complex relationship involving model, mimic, and receiver, in which S2 in the mimic is *better explained* by the receiver's tendency to treat S1 as an indicator of F in the model than by a direct relationship between S2 and F in the mimic. Mimicry arguments against the consciousness of AI systems rely, implicitly or explicitly, on the presence of this complex causal structure.

Mimicry, so conceived, is far more intricate than resemblance or imitation. A cloud may happen to look like a dog to some observer, but its dog-like shape is not explained by how observers interpret clouds. Likewise, if you follow my stone-hopping pattern across a stream simply to keep your feet dry, you imitate but do not mimic me. Mimicry always explains S2 via an intentional or functional relationship to a receiver's reaction – specifically, the receiver's reaction to the resemblance between S1 and S2 in light of the fact that S1 normally indicates F.

*8. Mimicry in Artificial Intelligence.*

Consider a child's toy that says "hello" when powered on. We can treat this as a simple case of mimicry. English-speaking humans are the models, the toy is the mimic, S1 and S2 are the sound "hello", F is the intention to greet, and the receiver is a child who hears the sound as an English "hello". In human models, "hello" normally (though not perfectly) indicates an intention to greet. But of course the toy has no intention to greet. (The toy's designer might have intended to craft a toy that would "greet" its user, but that isn't the *toy's* intention.) S2 (the toy's "hello") is best explained not as an indicator of F (a genuine intention to greet) but as an S2

---

(Sherratt 2008). Similarly, members of the local parrot group might reliably signal group membership by repeating the same calls. Honest mimicry differs from ordinary honest signaling because the bulk of the causal explanation of S2 depends on S1's rather than S2's relation to F. Such cases also suggest that mimicry can be a matter of degree and/or symmetric.

modeled on a S1 (a human "hello"), because of the anticipated reactions to S2 by receivers (children) who normally treat S1 as indicating F.

Large Language Models (LLMs) such as GPT, Gemini, LLaMA, and Claude are more complex but are structurally mimics, as Bender (Bender and Koller 2020; Bender et al. 2021), Millière (2022), and Jaeger (2023/2024) have argued. Our account renders this claim more precise. If you ask ChatGPT "What is the capital of California?", it might respond "The capital of California is Sacramento." Treating this output as S2, ChatGPT exhibits S2 because its outputs are modeled on human-produced text that also correctly identifies the capital of California as Sacramento, designed with a receiver in mind who will interpret that output linguistically. While human-produced text with that content (S1) reliably indicates the producer's knowledge that Sacramento is the capital (F), we cannot infer the presence of F from S2 – at least not without substantial (and in this case dubious) further argument.

Consider how transformer-based LLMs are pretrained through selection pressures. Patterns matching human text patterns are reinforced, entirely on the basis of that matching. Patterns failing to match human text patterns are suppressed, entirely on the basis of that failure. The only target of selection is similarity to human text. In pretraining, the model is not rewarded for tracking features of the outside world. Pure transformer models without post-training are designed so that their patterns of text output match those of ordinary human text, to a sufficient degree and for the sake of use by a human receiver. Although these patterns might correlate with, or indirectly track, features of the world beyond, the best and simplest explanation of a pure transformer's pretrained text outputs is in terms of their match to human-produced text outputs for the sake of a receiver who will interpret them as meaningful. If the human texts were different, the model's responses would be different, even if the world were the same. If the human texts were the same, the responses would be the same, even if the world were different. This is not an incidental shortcut. It is the fundamental nature of the selection pressure.[21] Mimicking patterns in human text is their core function and the best explanation of their behavior.

The situation becomes more complicated as the models become more complicated. In post-training, models can be rewarded for giving humanlike answers, which amplifies mimicry pressures. But they can also be rewarded for giving factually correct or desirable answers or punished for incorrect or undesirable answers, rewarded for acting successfully toward goals and punished for acting unsuccessfully. Some also incorporate elements like maps and calculators that have representational rather than mimicry functions. Consequently, they diverge from being *pure* mimics. They are no longer 100% "parrot" (or underground octopus).[22] However, in our current technological climate, LLMs' core functionality – the main reason they emit S2s that resemble our S1s and are interpretable by us – relies on mimicry structures created in pretraining. If someday we can train models primarily through direct engagement with the world itself, rewarding and punishing them for achieving real-world communicative or other goals, the mimicry argument will no longer apply.

Mimicry needn't always be *mere* mimicry. Feature F might be present, either unconnected with S2 or as a means of attaining S2. Imagine if the easiest way to achieve the wing coloration pattern of a monarch were to produce a chemical toxic to predators. Analogously, LLMs might sometimes succeed in mimicry by acquiring internal states that resemble those of humans.[23] The existence of mimicry doesn't preclude this possibility. It just shifts the explanatory burden. Given a mimicry explanation of S2 without appeal to F, to establish the existence of F requires either a direct argument for F's existence or an argument that S2 requires F.

---

[21] The case thus differs from the much discussed case of frogs eating BBs, where the individual organism tracks something like *small, dark, and moving* but the species-level selective pressure is for something like *nutritious food* (Neander 2017).

[22] Bender and Koller 2020; Bender et al. 2021.

*9. Consciousness Mimicry.*

We can call something a *consciousness mimic* if it exhibits readily observable features suggestive of consciousness whose presence is best explained by having been modeled on similar observable features of another system, for the sake of a receiver who treats those features, when they occur in the model, as indicating a conscious state or capacity.

Note that consciousness mimicry, so defined, does not require mimicking consciousness *per se*. Mimicking text patterns can count as consciousness mimicry if the receiver normally treats such patterns, when they occur in humans, as indicating consciousness. Nor need there be any attempt to mislead the receiver or to evoke responses receivers normally have to conscious entities. Compare: The mime who mimics the gait of a depressed person seeks to produce laughter in the receivers, not reactions appropriate to depression. Nor need they mimic any aspect of depression other than its gait.

ChatGPT and the "hello"-toy are consciousness mimics in this sense. Their observable language-like features (their S2s) are modeled on us, for receivers with certain interests and background knowledge. In us, the modeled features (the S1s) reliably indicate the further feature (F) of consciousness (humans who say "hello" are normally conscious), but there needn't be any such connection in a consciousness mimic. "Social AI" programs, like Replika, embellish the mimicry with an animated face, without changing the fundamental mimicry relation. In the "hello"-toy it's obvious that although S1 indicates consciousness in the model human, S2 need not do so in the mimic. The Mimicry Argument generalizes this observation to less obvious cases where mimicry is more extensive.

[23] For philosophical arguments that LLMs have humanlike cognitive or representational structures, see Cappelen and Dever 2025; Tye 2025.

Contrast such consciousness mimicry with children's language learning. Children learn through imitation, and sometimes mimicry, but the primary structure is not mimicry. If you say "blicket" in the presence of a blicket and the child repeats "blicket", their behavior arises from their (presumably conscious) understanding that the novel word refers to the novel object. Their S2 resembles your S1, for the sake of a receiver (you), but we explain S2 in terms of the child's own underlying F (knowledge that the object is a blicket). Children's crying, demands for food, and insistent *noes* are not S2s explained by S1s linked to Fs in a model. They are direct signals (normally) of genuine unhappiness, desire, and reluctance. Similarly, Alex the parrot's utterances of "four corner paper" in the presence of square pieces of paper – though drawing on the capacities that underlie parrot mimicry – function as true signals in the ordinary, direct sense (Pepperberg 1999). Childhood linguistic mimicry occurs, but it differs from ordinary language learning. One example would be in shared pretense, when a child says, "I've got to hurry or I'll be late to work!", theatrically grabbing a pretend briefcase.

*10. The Structure of the Mimicry Argument.*

The default in Default Liberalism about the consciousness of behaviorally sophisticated entities can thus be canceled in at least two ways: (a.) if the behavior or structure suggests a sufficiently important deficit, or (b.) if the entity is a consciousness mimic.

In particular, the Mimicry Argument can be expressed as follows:

(1.) In cases of mimicry, the appearance of S2 in the mimic is not normally sufficient grounds to infer the presence of F in the mimic.

(2.) An important range of AI systems are consciousness mimics.

(3.) Therefore, in this range of AI systems, we cannot infer consciousness from the (S2) features that, when their analogs (S1) to occur in humans, normally indicate conscious states (F).

As discussed, S2 mimicry traits aren't *incompatible* with the existence, in the mimic, of the F traits normally indicated by S1s in the model. A viceroy might be toxic; a parrot mimicking a group signal might belong (in fact, usually does belong) to the targeted group. But establishing that requires further evidence or argument.

The Mimicry Argument doesn't establish that consciousness mimics are definitely nonconscious. It only undercuts the inference from superficial appearance (S2) to underlying reality (F). Perhaps a compelling argument can be made for AI consciousness in some range of current or near-future cases; but given the degree of uncertainty in consciousness science, we doubt any such argument would warrant scientific and philosophical consensus.[24]

## *11. Are AI Systems "Behaviorally Sophisticated" in the Sense Relevant to the Copernican Principle?*

In Section 2, we defined a "behaviorally sophisticated" entity as one capable of complex goal-seeking, communication, and cooperation, where these terms are understood purely behaviorally, without presupposing consciousness or any particular cognitive architecture. Are current or near-future AI systems behaviorally sophisticated in this sense, and if so, does the Copernican Principle of Consciousness apply to them?

[24] For discussion, see Butlin et al. 2023; Schwitzgebel 2025/2026.

For alien cases, we favor a somewhat relaxed interpretation of "behavioral sophistication". As contemplated in Section 5, differently structured entities might display approximately human-level goal-seeking, communication, and cooperation despite deficits that we would find extraordinary (e.g., major deficits in learning or self-representation). We human beings also have remarkable cognitive weaknesses. We wouldn't want an alien species to judge us behaviorally unsophisticated because of our – possibly to them shocking – weakness in formal logical reasoning. Entities with different cognitive structures will likely have different strengths and weaknesses. A proper Copernican perspective won't insist that alien intelligence look too much like ours.

Applying this same liberality to AI cases, we don't insist that present and near-future AI systems lack approximately human-level behavioral sophistication. On some tasks – such as walking across uneven surfaces and identifying objects in blurry, partly occluded photographs – human beings (so far) substantially exceed AI systems. On other tasks – such as playing chess and multiplying large numbers – AI systems substantially exceed even expert humans. It might be tempting for opponents of AI consciousness to rest their case on deficits in some crucial range of tasks. However, this requires correctly specifying the relevant tasks, a notoriously slippery project. AI enthusiasts justifiably object to a history of moving goalposts. Back when chess seemed like the pinnacle of human expertise, AI skeptics sometimes treated chess as a special sign of human intelligence. The Turing-inspired emphasis on the ability to hold humanlike conversation also increasingly looks like a poor indicator, or at least one requiring high and precise standards that are potentially difficult to justify. The Mimicry Argument allows AI skeptics to express skepticism about AI consciousness without (a.) denying that AI might match or surpass humans in behavioral sophistication overall, (b.) relying on a privileged behavioral benchmark that AI cannot pass, or (c.) committing to a specific theory of consciousness or specific claims about the types of architectures than can and cannot support consciousness. These are, we submit, attractive features of the Mimicry Argument.

If we concede that mimicry-based AI systems are behaviorally sophisticated, does that undercut the Copernican Argument? It does if it prevents us from treating behaviorally sophisticated entities as the reference class for the Copernican Principle of Consciousness. If we know that we are conscious and (let's suppose) that current mimicry-based AI systems are not, then we might have grounds to regard ourselves as special among the class of behaviorally sophisticated entities, much as we have grounds to regard ourselves as special among the class of carbon containing entities.

We see three responses. One is to assume that behaviorally sophisticated consciousness mimics are overall rare in the universe (see Section 12). There would then be no reason to regard ourselves as special among the class of behaviorally sophisticated entities after all. Another response modifies the reference class of the Copernican Principle to include only behaviorally sophisticated non-mimics. A third response endorses agnosticism about AI consciousness. The Mimicry Argument undercuts the inference from superficial behavior to underlying consciousness, justifying skepticism. But only with the help of other assumptions, which we don't defend here, can it deliver the definite conclusion that behaviorally sophisticated AI are *not* conscious.

## *12. Consciousness Mimics and Alien Mimics.*

Could the pressures on a consciousness mimic lead it to develop genuine consciousness? The answer depends on the receiver's interests and sophistication. If selection pressures are severe and specific enough to generate complex goal-seeking, communication, and cooperation, those same pressures might also produce complex cognitive structures of the sort practically required to sustain such behavior – for example, long-term memory, self-representation, and linguistic understanding. Fantastical constructs, such as "Blockheads" (Block 1981) might be behaviorally identical to a naturally selected intelligent organism, but they could never evolve: The obvious way to reliably act, over evolutionary time, as if one has a long-term memory is to actually have a long-term memory. In contrast, a consciousness mimic's behavioral sophistication need only suffice for the goals of mimicry – for example, to fool or delight a receiver. We are often delighted by obviously-less-than-perfect mimics (the "hello"-toy) and, as recent history suggests, LLM mimics sometimes easily fool us.

Different types and degrees of similarity between S2 and S1 are required for different mimicry relationships. Deceptive mimicry can become an arms race between a mimic motivated to appear similar to the model in the eyes of the intended dupe and an intended dupe motivated to acquire the capacities to distinguish mimic from model. In the limit, the mimic could only "fool" a god-like receiver by actually acquiring feature F. Fortunately for mimics, most receivers are far from god-like.

There's a reason most features of most evolved organisms aren't explained by mimicry. Mimicry is a complex relationship requiring a special set of evolutionary pressures, often parasitic on the true signals or cues of more prevalent non-mimic species. If viceroys outnumber monarchs, the evolutionary advantage of mimicry decreases. Technological mimicry might also prove transient. In the long run, human design goals might be met more fully and efficiently by creating systems that really have F than by creating systems that imitate the superficial signs of F. An artificial entity that genuinely understands itself and its interlocutors (whatever architecture that requires) will presumably be a more reliable, versatile, and satisfying interaction partner than today's Large Language Models. Similarly, if our extraterrestrial neighbors dispatch robotic probes across the galaxy, it's reasonable to expect the probes will be designed to explore, measure, experiment, communicate, self-repair, respond to emergencies, and perhaps devise theories on the fly that guide their exploratory choices, rather than designed as mimics to fool or delight some receiver through their resemblance to some model entity.

Since mimicry traits evolve less commonly than non-mimicry traits and require highly specific selective or design conditions, it is reasonable to assume – tentatively and as a default – that newly encountered behaviorally sophisticated extraterrestrial entities will not be consciousness mimics, even if they have a technological origin.

*13. Rejecting the Parity Principle.*

We sympathize with those who reject a simple sort of organic chauvinism – the idea that only biological entities can be conscious. The most obvious alternative is to treat alien and robot cases symmetrically – holding that the same behavioral and architectural criteria apply to both when assessing them for consciousness. On this view, if Behavioral Test X is sufficient to warrant attribution of alien consciousness, then we should attribute consciousness to any robot passing the same behavioral test. Similarly, if Architecture Y would warrant attribution of consciousness to aliens, it should also suffice for robots. According to the *Parity Principle,* we should apply the same behavioral and/or architectural criteria in evaluating the consciousness of aliens and robots.

The Parity Principle has intuitive appeal. But to accurately assess it, we need to distinguish it from a related metaphysical thesis. Metaphysically, if an alien and a robot are identical in everything that matters for consciousness according to the correct theory of consciousness (whatever that may be), then either both or neither are consciousness. This metaphysical parity thesis is highly plausible.

The Parity Principle we reject differs from the plausible metaphysical thesis in two respects. First, our Parity Principle is epistemic, not metaphysical. It concerns not what must be true if robots and aliens share whatever behavioral or architectural features matter to consciousness; rather it concerns what we are *warranted in believing* about alien and robot consciousness *given our ignorance* about the correct theory of consciousness. Second, it applies without requiring stipulations about the entities' internal workings (even if it can be undercut by stipulating unpromising enough internal workings: Section 5).

These differences matter. Even if we know virtually nothing about the internal workings of a particular (individual or type of) behaviorally intelligent alien, we can apply the Copernican Principle by default (that is, assuming we have no evidence it is a consciousness mimic or in some other respect likely to be a member of a disprivileged class with respect to consciousness). Even if we know virtually nothing about the internal workings of a particular (individual or type of) robot or AI system, if we know that the superficial features suggestive of consciousness arise from a mimicry relationship of the sort described in Section 7, we are licensed to withhold inference to consciousness, barring positive evidence that the superficial features are coupled with consciousness.

The Copernican and Mimicry Arguments yield different results for aliens and robots not because of some intrinsic inferiority of engineered to naturally evolved life but rather because aliens and a wide range of current and near-future robots – normally or as a default presumption – have very different *histories*. Consequently, very different explanations of their structure and behavior can be warranted. Contra the Parity Thesis, this difference generates an epistemic asymmetry in attribution of consciousness to aliens and robots, even if the aliens and robots have similar outward behavior – or even similar interior architectures, at a coarse level of descriptive specificity. (However, if the interior architectures are sufficiently similar, metaphysical parity might again become plausible.)

This is not to endorse the teleofunctionism of Millikan (1984), Dretske (1995), and Neander (2017), according to which Donald Davidson has a mind and his molecule-for-molecule identical chance-generated Swampman duplicate lacks a mind because of their different origins. Such teleofunctionalism concerns the metaphysical relationship of mental properties to their physical realizers at a given time. Our approach is neutral about the metaphysical importance of an entity's history while remaining committed to the history's epistemic importance. History matters for what we are *justified in believing* even if it doesn't *metaphysically determine* whether consciousness is present.

*14. Generalizing the Copernican Argument to Less Sophisticated Aliens.*

In Section 2, we employed a high bar for "behavioral sophistication" because we were seeking a sufficiency criterion for default attributions of alien consciousness – one that would appeal across the theoretical spectrum from very liberal views (on which consciousness is widespread on Earth) to very conservative views (on which consciousness is rare). But this is a flexible parameter. The Copernican Argument can be generalized to non-human animal cases as follows.

Barring some reason to think Earth is exceptional, we would be strangely lucky to inhabit a planet populated with conscious non-linguistic life forms while, almost everywhere else, similarly complex or sophisticated life forms are nonconscious. Whatever types of life you are justified in regarding as conscious on Earth (and different readers might reasonably favor different views), it is reasonable to assume that similarly complex or sophisticated alien life forms would also be conscious.[25]

[25] For helpful discussion, thanks to Brice Bantegnie, Cameron Buckner, Nathan Buss, David Chalmers, Leonard Dung, Kenny Easwaran, Riley Harris, Caleb Hazelwood, Benjamin Icard, Andrew Lee, Matthew Liao, Michael Martin, Raphaël Millière, Saam Nasseri Mashhadi, Melissa O'Neill, Jonathan Pober, William Robinson, Nick Shea, Stephan Schmid, Eddie Schwieterman, and Anna Strasser; audiences at Trent University, Harvey Mudd, New York University, the Agency and Intentions in AI conference in Göttingen, Jagiellonian University, the Oxford Mind Seminar, University of Lisbon, NOVA Lisbon University, University of Hamburg, the Philosophy of Neuroscience/Mind Writing Group, Duke University, and Eric Schwitzgebel's Winter 2025 graduate seminar on alien and AI consciousness; and commenters on relevant posts on social media and at The Splintered Mind.

<u>References:</u>